\documentclass[11pt,titlepage,a4paper]{article}
\usepackage{bozhomac}  
\usepackage{bozhlogo}  
\usepackage{cite}	

\title{\bfseries	\vspace*{-1.678902345in}
{\huge On momentum operator\\[1ex] in quantum field theory}
}

\vspace{1.7ex}

\author{
Bozhidar Z.\ Iliev
\thanks{Department of Mathematical Modeling,
Institute for Nuclear Research and \mbox{Nuclear} Energy,
Bulgarian Academy of Sciences,
Boul.\ Tzarigradsko chauss\'ee~72, 1784 Sofia, Bulgaria}
\thanks{E-mail address: bozho@inrne.bas.bg}
\thanks{URL: http://theo.inrne.bas.bg/$\sim$bozho/}
}

\date{	
 \vspace{2.27ex}\ShortTitle{Momentum operator in QFT}\\[0.27ex]
 \vspace{3.27ex}
\small
	\begin{tabular}{r@{$\colon\to~$}l}
 \vspace{0.09ex} Basic ideas	& March -June, 2001		\\[0.09ex]
 \vspace{0.09ex} Began		& March 20, 2001	\\[0.09ex]
 \vspace{0.09ex} Ended		& June 21, 2001 	\\[0.09ex]
 \vspace{0.09ex} Initial typeset& March 28, 2001--June 25, 2001	\\[0.09ex]
\vspace{0.09ex} Last update	& June 2, 2002\\[0.09ex]
 \vspace{0.27ex} Produced	& \fbox{\today}	\\[0.27ex]
	\end{tabular} \\[1.27ex]
\normalsize
	\begin{tabular}{r@{$\colon~$}l}
\vspace{0.27ex} LANL xxx archive server E-print No. & hep-th/0206008
	\end{tabular} \\[-0.27ex]
 \vspace{4.27ex}{\Huge\BOZHO}	\\[4.27ex]
 \vspace{0.27ex}\Subject{Quantum field theory}
								\\[2.27ex]
	\begin{tabular}{r@{\hspace{0.512em}}|@{\hspace{0.512em}}l}
 \vspace{0.27ex}\MSC[2000]{81Q99,81T99\\\hspace{0pt}}
&
 \vspace{0.27ex}
	\end{tabular} \\[1.27ex]
 \vspace{0.27ex}\\[0.27ex]
}

\newcommand{\Hil}{\mathcal{F}}		
	\newcommand{\base}{\mathit{M}}	

 \newcommand{\Ham}{\mathcal{H}}	
 \newcommand{\dyn}[1]{\pmb{\mathbb{#1}}}	

\newcommand{\ope}[2][{}]{\lindex[\mathcal{#2}]{}{#1}} 
\begin{document}		

\renewcommand{\thefootnote}{\fnsymbol{footnote}} 
\maketitle				
\renewcommand{\thefootnote}{\arabic{footnote}}   

\tableofcontents		


	\begin{abstract}
	The interrelations between the two definitions of momentum operator,
via the canonical energy\ndash momentum tensorial operator and as translation
operator (on the operator space), are studied in quantum field theory. These
definitions give rise to similar but, generally, different momentum
operators, each of them having its own place in the theory. Some speculations
on the relations between quantum field theory and quantum mechanics are
presented.
	\end{abstract}

\section {Introduction}
\label{Introduction}

	Two definitions of a momentum operator exist in quantum field theory.
The first one defines it as a conserved operator arising via the Noether's
theorem for translation invariant Lagrangians; we call the arising operator
the canonical (or physical) momentum operator, or simply the momentum
operator. The second definition defines  the momentum operator as a generator
of the representation of translations in the Minkowski spacetime on the space
of operators acting on the Hilbert space of some system of quantum fields;
we call the so\ndash arising operator the translation (or mathematical)
momentum operator. As we shall see, this second operator is defined up to a
constant 4\ndash vector which allows its identification with the physical
momentum operator on some subset of the Hilbert space of states of a quantum
system; as a rule, that subset is a proper subset.

	The lay-out of the work is as follows.

	In Sect.~\ref{Sect1} the rigorous formulation of the definitions
mentioned above are presented and a relation is derived between the
(physical) momentum operator and the translation operator acting on system's
Hilbert space; the last operator being identical with the momentum operator
used in quantum mechanics. In Sect.~\ref{Sect2} an analysis of these
definitions is done and some relations between the different momentum
operators are established. Certain conclusions from these results are made
in Sect.~\ref{Sect3}. In particular, the possible equivalence between
physical and mathematical momentum operators is discussed and the inference
is made that, generally, these are different operators. The results obtained
are summarized  in Sect.~\ref{Conclusion}. It also contains some speculations
on the links between quantum field theory and (non\ndash relativistic and
relativistic) quantum mechanics.

	In what follows, we suppose that there is given a system of quantum
fields, described via field operators $\varphi_i(x)$,
$i=1,\dots,n\in\field[N]$, $x\in\base$ over the 4\ndash dimensional Minkowski
spacetime $\base$ endowed with standard Lorentzian metric tensor
$\eta_{\mu\nu}$ with signature $(+\,-\,-\,-)$. The system's Hilbert space of
states is denoted by $\Hil$ and all considerations are in Heisenberg picture
of motion if the opposite is not stated explicitly. The Greek indices
$\mu,\nu,\dots$ run from 0 to $3=\dim\base-1$ and the Einstein's summation
convention is assumed over indices repeating on different levels. The
coordinates of a point $x\in\base$ are denoted by $x^\mu$, $\bs
x:=(x^1,x^2,x^3)$ and the derivative with respect to $x^\mu$ is
$\frac{\pd}{\pd x^\mu}=:\pd_\mu$. The imaginary unit is denoted by $\iu$ and
$\hbar$ and $c$ stand for the Planck's constant (divided by $2\pi$) and the
velocity of light in vacuum, respectively.


\section{The two momentum operators}
	\label{Sect1}

	There are two quite similar understandings of the energy-momentum
vectorial operator, called simply (4\ndash)momentum operator and denoted by
$\ope{P}_\mu$, in (translation invariant) quantum field theory.

	The first definition of momentum operator defines it, in the
Lagrangian formalism, through the canonical energy\ndash momentum tensorial
operator $\ope{T}_{\mu\nu}$~\cite[eq.~(2-41)]{Roman-QFT} of a system of
quantum fields $\varphi_i(x)$, \ie%
\footnote{~%
See, e.g.,~\cite{Bogolyubov&Shirkov,Roman-QFT,Bjorken&Drell,Itzykson&Zuber}.%
}
	\begin{equation}	\label{6.1}
\ope{P}_\mu
:=
\frac{1}{c} \int \Sprindex[\ope{T}]{\mu}{0}(x) \Id^3\bs x .
	\end{equation}
Here the integration is over the equal-time surface $x^0=ct=\const$ with $c$
being the velocity of light, $x$ is a point in the Minkowski spacetime
$\base$ of special relativity endowed with a metric tensor $\eta_{\mu\nu}$,
$[\eta_{\mu\nu}]=\diag(1,-1,-1,-1)$ by means of which are raised/lowered the
Greek indices.
We call the so\ndash defined operator $\ope{P}_\mu$ the canonical (or
physical) momentum operator or simply the \emph{momentum operator}. The
operator~\eref{6.1} is a conserved
quantity~\cite{Bogolyubov&Shirkov,Roman-QFT,Bjorken&Drell}, \ie
$\frac{\od\ope{P}_\mu}{\od x^0}=0$. So, since $\ope{P}_\mu$ does not depend
on $x^a$ for $a=1,2,3$, it is valid the equality
	\begin{equation}	\label{6.1new}
\frac{\pd\ope{P}_\mu}{\pd x^\nu}=0 .
	\end{equation}

	The second definition of the momentum operator identifies it with
the generator of representation of the translation subgroup of the Poincar\'e
group on the space of operators acting on the Hilbert space $\Hil$ of some
system of quantum fields $\varphi_i(x)$.%
\footnote{~%
See, for instance,~\cite[p.~146]{Bogolyubov&et_al.-AxQFT},
\cite[sec.~7.1]{Bogolyubov&et_al.-QFT},
\cite[sec.~2.1.1~(iiia)]{Roman-QFT},
\cite[sec.~3.1.2]{Itzykson&Zuber},
and~\cite[\S~68]{Bjorken&Drell-2}.%
}
The so\ndash arising operator will be referred as the \emph{translation} (or
mathematical) \emph{momentum operator} and will be denoted by
$\ope{P}_\mu^{\mathrm{t}}$.  It is defined as follows. Let $a$ and $x$ be
points in Minkowski spacetime $\base$ and $\ope{A}(x)\colon\Hil\to\Hil$ be a
field operator $\varphi_i(x)$ or an operator corresponding to some dynamical
variable $\dyn{A}$. The translation $x\mapsto x+a$ in $\base$ entails
	\begin{gather}	\label{6.2}
\ope{A}(x)\mapsto \ope{A}(x+a)
=
\e^{-\iih a^\mu \ope{P}_\mu^{\mathrm{t}}}
\circ\ope{A}(x)\circ
\e^{\iih a^\mu \ope{P}_\mu^{\mathrm{t}}},
\intertext{where $\hbar$ is the Planck constant (divided by $2\pi$), which, in
a differential form, can be rewritten as}
			\tag{\ref{6.2}$^\prime$}	\label{6.2'}
\ih\frac{\pd}{\pd x^\mu} \ope{A}(x)
=
[\ope{A}(x), \ope{P}_\mu^{\mathrm{t}} ]_{\_}
	\end{gather}
where  $\circ$ denotes the composition of mappings and
$[\ope{A},\ope{B}]_{\_}:=\ope{A}\circ\ope{B}-\ope{B}\circ\ope{A}$
is the commutator of $\ope{A},\ope{B}\colon\Hil\to\Hil$.

	There is a simple relation between $\ope{P}_\mu^{\mathrm{t}}$ and
the translation operator on $\Hil$. Let $\ope{P}_\mu^{\mathrm{QM}}$, where
the superscript QM stands for Quantum Mechanics (see Sect.~\ref{Conclusion}),
denotes the generator of translation operator on system's Hilbert space
$\Hil$, \ie $\ope{P}_\mu^{\mathrm{QM}}$ is the Hermitian generator of the
mapping $\ope{X}(x)\mapsto\ope{X}(x+a)$ for any points $x,a\in\base$.
Explicitly, we have
	\begin{equation}	\label{6.3}
\ope{X}(x)\mapsto \ope{X}(x+a)
=
\e^{\iih a^\mu \ope{P}_\mu^{\mathrm{QM}}}  ( \ope{X}(x) ) .
	\end{equation}

	The equality
	\begin{equation}	\label{2.1}
\ope{A}(x+a)
=
\e^{\iih a^\mu \ope{P}_\mu^{\mathrm{QM}}}
\circ\ope{A}(x)\circ
\e^{-\iih a^\mu \ope{P}_\mu^{\mathrm{QM}}}
	\end{equation}
is a simple corollary of~\eref{6.3}:
\(
\e^{+\dots} \circ\ope{A}(x)\circ \e^{-\dots} (\ope{X}(x))
=
\e^{+\dots} (\ope{A}(x) (\ope{X}(x-a)) )
=
\ope{A}(x+a) (\ope{X}(x-a+a))
=
\ope{A}(x+a) (\ope{X}(x)) .
\)
Similarly to~\eref{6.2'}, the differential form of~\eref{2.1} is
	\begin{equation}
			\tag{\ref{2.1}$^\prime$}	\label{2.1'}
\ih\frac{\pd}{\pd x^\mu} \ope{A}(x)
=
- [\ope{A}(x), \ope{P}_\mu^{\mathrm{QM}} ]_{\_} .
	\end{equation}

	Subtracting~\eref{6.2'} from~\eref{2.1'}, we find
	\begin{equation}	\label{2.3}
[ \ope{A}(x), \ope{P}_\mu^{\mathrm{QM}} + \ope{P}_\mu^{\mathrm{t}} ]_{\_}
= 0 .
	\end{equation}
Consequently, if $\ope{A}$ is \emph{arbitrary}, this equality, by virtue of
Schur's lemma%
\footnote{~%
See, e.g,~\cite[appendix~II]{Rumer&Fet}, \cite[sec.~8.2]{Kirillov-1976},
\cite[ch.~5, sec.~3]{Barut&Roczka}.%
}
implies
	\begin{equation}	\label{2.4}
\ope{P}_\mu^{\mathrm{t}} = - \ope{P}_\mu^{\mathrm{QM}} + p_\mu \id_\Hil
	\end{equation}
where $\id_\Hil$ is the identity mapping of $\Hil$ and $p_\mu$ are constant
(real --- see Sect.~\ref{Sect3}) numbers, with dimension of 4\ndash momentum,
representing the covariant components of some vector $p$.

	Notice (cf.~\cite[subsec.~9.3]{Bogolyubov&Shirkov}, if
$\langle \cdot|\cdot \rangle\colon\Hil\to\Hil$ is the Hermitian scaler
product of $\Hil$, the definition~\eref{6.2} is chosen so that
\(
\langle\ope{X}(x) |\ope{A}(x+a)(\ope{Y}(x)) \rangle
=
\langle\ope{X}(x+a) |\ope{A}(x)(\ope{Y}(x+a)) \rangle,
\)
while in a case of~\eref{2.1} it is fulfilled
\(
\langle\ope{X}(x+a) |\ope{A}(x+a)(\ope{Y}(x+a)) \rangle
=
\langle\ope{X}(x) |\ope{A}(x)(\ope{Y}(x)) \rangle
\)
for any $\ope{X}(x),\ope{Y}(x)\in\Hil$.


\section {Discussion}
\label{Sect2}

	At this point a problem arises: are the operators
 $\ope{P}_\mu^{\mathrm{t}}$ and $\ope{P}_\mu$ identical?
That is, can we write
\(
\ope{P}_\mu^{\mathrm{t}} = \ope{P}_\mu
\)
and identify this operator with system's `true' momentum operator,
or, more generally, what is the connection between
 $\ope{P}_\mu^{\mathrm{t}}$ and  $\ope{P}_\mu$, if any?
The author of these lines fails to find the answer, as well as the explicit
formulation, of that problem in the literature; the only exception
being~\cite[\S~68]{Bjorken&Drell-2} where in the discussion following
eq.~(11.71) in \emph{loc.\ cit.}, it is mentioned of a possible problem that
may arise if the canonical momentum operator, in our terminology, does not
satisfy equations like~\eref{6.8} below.

	Consider two approaches to the above problem.

	In~\cite[subsec.~2.1, p.~70]{Roman-QFT} is \emph{implicitly} assumed a
connection between
 $\ope{P}_\mu^{\mathrm{t}}$ and $\ope{P}_\mu$,
by, in fact, postulating that~\eref{6.2'} should be valid for
$\ope{P}_\mu$,%
\footnote{\label{CounterExample}~
In fact, in~\cite[subsec.~2.1]{Roman-QFT} is actually \emph{proved}
equation~\eref{6.8} below from which follows~\eref{6.4} for $\ope{A}(x)$
of a type of polynomial or convergent power series in the  field operators
$\varphi_i(x)$. Actually the equality~\eref{6.4} cannot hold for arbitrary
operator $\ope{A}(x)$. A simple counterexample is provided by any quantum
system possessing a non\ndash vanishing momentum and angular momentum
operators $\ope{P}_{\mu}$ and $\ope{M}_{\mu\nu}$, respectively. In deed, on
one hand, we have $\pd_\lambda\ope{M}_{\mu\nu}=0$ as $\ope{M}_{\mu\nu}$ is
a conserved quantity~\cite{Roman-QFT,Bogolyubov&Shirkov,Bjorken&Drell-2},
and, on other hand,
\(
[ \ope{M}_{\mu\nu} , \ope{P}_{\lambda} ]_{\_}
=
\ih ( \eta_{\lambda\mu} \ope{P}_{\nu} - \eta_{\lambda\nu} \ope{P}_{\mu} )
\)
(see, e.g.,~\cite[p.~77]{Roman-QFT}) and, consequently
\(
\ih\frac{\pd\ope{M}_{\mu\nu}}{\pd x^\lambda}
= 0
\not=
[ \ope{M}_{\mu\nu} , \ope{P}_{\lambda} ]_{\_} .
\)%
} 
 \ie
	\begin{equation}	\label{6.4}
\ih\frac{\pd\ope{A}(x)}{\pd x^\mu}
=
[ \ope{A}(x), \ope{P}_\mu ]_{\_}
	\end{equation}
for any operator $\ope{A}(x)$. Combining this with~\eref{2.1'}, we get
\[
[ \ope{A}(x) , \ope{P}_\mu - \ope{P}_\mu^{\mathrm{t}} ]_{\_} = 0
\]
for \emph{every} operator $\ope{A}\colon\Hil\to\Hil$. Hereof the Schur's
lemma implies
	\begin{equation}	\label{6.5}
\ope{P}_\mu =  \ope{P}_\mu^{\mathrm{t}} + q_\mu \id_\Hil
	\end{equation}
for some 4-vector field $q$ with \emph{constant} covariant components $q_\mu$.

	In~\cite[\S\S~9,10]{Bogolyubov&Shirkov}, we see a mixture of
 $\ope{P}_\mu$ and $\ope{P}_\mu^{\mathrm{t}}$,
denoted there by the single symbol $P^n$, and only with some effort, from the
context, one can find out whether the authors have in mind
 $\ope{P}_\mu$ or  $\ope{P}_\mu^{\mathrm{t}}$.
It is known, the generators of translations in a vector space $\Hil$
are the partial derivative operators $\pd_\mu$ on $\Hil$, so that the
explicit form of $\ope{P}_\mu^{\mathrm{QM}}$ is%
\footnote{~%
Proof: Expanding $\ope{X}(x+a)=\sum_{\sigma}\ope{X}^\sigma e_\sigma$, where
$\{e_\sigma\}$ is an independent of $x$ basis of $\Hil$,
into a Taylor's series around the point $x$, we get
	\begin{multline*}
\ope{X}(x+a)
 =
\sum_{\sigma} \Bigl( \sum_{n=0}^{\infty}\frac{1}{n!}
	a^{\mu_1}\cdots a^{\mu_n}
	\frac{\pd^n \ope{X}^\sigma}{\pd x^{\mu_1}\dots\pd x^{\mu_n}} \Big|_{x}
\Bigr) e_\sigma
 =
\sum_{\sigma} \Bigl( \sum_{n=0}^{\infty}\frac{1}{n!}
	(a^{\mu_1}\pd_{\mu_1})
\cdots (a^{\mu_n}\pd_{\mu_n}) \ope{X}^\sigma
\Bigr)\Big|_{x} e_\sigma
\\ =
\sum_{\sigma} \Bigl( \sum_{n=0}^{\infty}\frac{1}{n!}
	[ (a^{\mu}\pd_{\mu})^n \ope{X}^\sigma ]\Big|_{x}
\Bigr) e_\sigma
 =
\sum_{n=0}^{\infty}\frac{1}{n!}
	[ (a^{\mu}\pd_{\mu}^{\Hil})^n \ope{X} ]\Big|_{x}
=
\bigl( \e^{a^{\mu}\pd_{\mu}^{\Hil}} (\ope{X})\bigr) (x),
	\end{multline*}
that is
\(
\e^{a^{\mu}\pd_{\mu}^{\Hil}} \colon \ope{X}
\mapsto
\e^{a^{\mu}\pd_{\mu}^{\Hil}} (\ope{X}) \colon x
\mapsto
\ope{X}(x+a).
\)
Using some freedom of the notation, one usually writes
$\frac{\pd}{\pd x^\mu}$  for $\pd_{\mu}^{\Hil}$ and the last result is
written in the form~\eref{6.3}.%
}
	\begin{equation}	\label{6.6}
\ope{P}_\mu^{\mathrm{QM}} = \ih \pd_\mu^\Hil ,
	\end{equation}
where, if $\ope{X}\colon x\mapsto\ope{X}(x)\in\Hil$,
\(
\pd_\mu^\Hil \colon\ope{X}\mapsto
\pd_\mu^\Hil(\ope{X}) \colon x \mapsto
\frac{\pd\ope{X}}{\pd x^\mu}\big|_x.
\)
This equality is practically derived
in~\cite[subsec.~9.3]{Bogolyubov&Shirkov} where a remark is made that the
\emph{r.h.s.\ of~\eref{6.6} cannot serve as an energy\ndash momentum
vectorial operator}
as its application to any state vector, which is constant in
the Heisenberg picture, is identically equal to zero. The conclusion is that
the momentum operator must be something else, specified in the \emph{loc.\
cit.}\ as our translation momentum operator $\ope{P}_\mu^{\mathrm{QM}}$
until subsection~10.2 in \emph{loc.\ cit.}, where as it is taken our physical
momentum operator $\ope{P}_\mu$. However, the final results
in~\cite{Bogolyubov&Shirkov} are correct as, in fact, in \emph{loc.\ cit.}\
only the relations (cf.~\eref{6.2'})
	\begin{equation}	\label{6.7}
\ih\frac{\pd}{\pd x^\mu} \varphi_i(x)
=
[\varphi_i(x), \ope{P}_\mu^{\mathrm{t}} ]_{\_} ,
	\end{equation}
where $\varphi_i(x)$ is any field operator, are employed (see, e.g.,
subsections~9.3--10.1 in \emph{loc.\ cit.}). Further,
in~\cite[subsec.~10.2]{Bogolyubov&Shirkov}, the authors assume~\eref{6.7} to
hold with
 $\ope{P}_\mu$ for  $\ope{P}_\mu^{\mathrm{t}}$,%
\footnote{~%
This is a theorem; see, e.g.,~\cite[p.~70]{Roman-QFT}. The commutation
relations must be compatible with~\eref{6.8}~\cite[\S~68]{Bjorken&Drell-2}.%
}
\ie (cf.~\eref{6.4})
	\begin{equation}	\label{6.8}
\ih\frac{\pd}{\pd x^\mu} \varphi_i(x)
=
[ 	\varphi_i(x), \ope{P}_\mu ]_{\_} .
	\end{equation}
From these equalities, know as \emph{Heisenberg relations or Heisenberg
equations of motion} for the field operators, they derive the
(anti)commutation relations for the frequency
parts of the field operators as well as for creation and annihilation
operators (for free fields). Consequently, the theory is so built that in it
both relations,~\eref{6.8} and~\eref{6.7}, hold.

	\begin{Rem}	\label{Rem2.1}
	The relations~\eref{6.8} are external for the Lagrangian formalism
and their validity depends on the particular Lagrangian employed.
Usually~\cite[\S~68]{Bjorken&Drell-2} only Lagrangians for which~\eref{6.8}
holds are used.
	\end{Rem}

	Suppose $\ope{A}(x)$ is function of the field operators and their
partial derivatives (of finite order). If $\ope{A}(x)$ does not
depend explicitly on $x$ as a separate argument and it is a polynomial or
convergent power series in the field operators and their derivatives,
then~\eref{6.8} implies~\eref{6.4} for such
functions $\ope{A}(x)$.%
\footnote{~%
To prove this, one has to calculate the commutators of every term in the
expansion of $\ope{A}(x)$ with $\ope{P}_\mu$ by using the identity
$[a\circ b,c]_{\_}\equiv[a,c]_{\_}\circ b + a\circ[b,c]_{\_}$
for any operators $a$, $b$, and $c$, to express the remaining commutators
through~\eref{6.8}, and, at the end, to sum all of the terms obtained. If
$\ope{A}(x)$ depends on derivatives of $\varphi_i(x)$, in the proof one
should take into account the equality
\(
[ \pd_\mu(\varphi_i) , \ope{P}_{\nu} ]_{\_}
=
\pd_\mu[ \varphi_i , \ope{P}_{\nu} ]_{\_}
=\ih \pd_\mu\circ\pd_\nu (\varphi_i),
\)
which is a corollary of $\pd_\mu\ope{P}_{\nu}\equiv0$.%
}
If we assume that such operators $\ope{A}(x)$ can form an irreducible unitary
represention of some group, then, equation~\eref{6.5} follows
from~\eref{6.4} and~\eref{6.2'} and the Schur's lemma.

	Similar is the situation in other works devoted to the grounds of
quantum field theory. E.g., in~\cite[\S~3.1.2]{Itzykson&Zuber} both momentum
operators are identified, while in~\cite[sec.~3.1]{Streater&Wightman} the
momentum operator is identified with the Hermitian generator of a unitary
representation of the translation subgroup of the Poincar\'e group.  However,
as we demonstrated, the generators of translations on the operator space are
defined up to a multiples of the identity mapping of the system's Hilbert
space. This arbitrariness seems not be used until now by implicitly setting
the mentioned multiplier to zero.


\section {Inferences}
\label{Sect3}

	In Sect.~\ref{Sect2} we saw that the translation momentum operator
$\ope{P}_\mu^{\mathrm{t}}$ of a system of quantum fields is given by the
r.h.s.\ of~\eref{2.4},
	\begin{equation}	\label{6.9}
\ope{P}_\mu^{\mathrm{t}} = -\ope{P}_\mu^{\mathrm{QM}} + p_\mu\id_\Hil ,
	\end{equation}
which, in view of~\eref{6.6}, may be rewritten as
	\begin{equation}	\label{6.10}
\ope{P}_\mu^{\mathrm{t}} = -\ih\frac{\pd}{\pd x^\mu} + p_\mu\id_\Hil ,
	\end{equation}
where $p_\mu$ are the covariant components of a 4-vector.%
\footnote{~%
Since the Lorentz/Poincar\'e transformations, employed in quantum field
theory, are linear transformations with \emph{constant} (in spacetime)
coefficients, the assertion that a vector has constant components is a
covariant one.%
}

It, obviously, satisfies the relation%
\footnote{~%
In fact, the r.h.s.\ of~\eref{6.9} is the general solution of~\eref{6.11}
with respect to $\ope{P}_\mu^{\mathrm{t}}$ if~\eref{6.11} holds for every
$\ope{A}(x)\colon\Hil\to\Hil$.%
}
	\begin{equation}	\label{6.11}
\ih\frac{\pd\ope{A}(x)}{\pd x^\mu}
=
[\ope{A}(x),\ope{P}_\mu^{\mathrm{t}}]_{\_}
	\end{equation}
for any $\ope{A}(x)\colon\Hil\to\Hil$. A little below (see the discussion
after~\eref{6.13}), the \emph{possible} equality
	\begin{equation}	\label{6.12}
\ope{P}_\mu^{\mathrm{t}} = \ope{P}_\mu
	\end{equation}
will be explored. But when results involving $\ope{P}_\mu^{\mathrm{t}}$ are
extracted from~\eref{6.11} in which $\ope{A}(x)$ is  (observable or not)
operator constructed form the field operators and their partial derivatives,
one can assumed~\eref{6.12} to hold until~\eref{6.11} with  $\ope{A}(x)$ from
a different kind is required. However, if we want to look on~\eref{6.11} as
(implicit Heisenberg) equations of motion, as it is done often in the
literature, the equality~\eref{6.12} is unacceptable since~\eref{6.11}
entails~\eref{6.4} as an identity. Consequently, when the equations of motion
have to be considered, one should work with~\eref{6.4}, not with~\eref{6.11}.
In particular, the \emph{Heisenberg equations of motion for the field
operators $\varphi_i$ are~\eref{6.8}}, not~\eref{6.7} or~\eref{6.11} with
$\ope{A}(x)=\varphi_i(x)$ and $\ope{P}_\mu^{\mathrm{t}}$ given by~\eref{6.9}.

	Assume now $\ope{X}\in\Hil$ is a state vector of a system of quantum
fields. It is a constant vector as we work in the Heisenberg picture, \ie we
have the equivalent equations
	\begin{equation}	\label{6.12new}
\e^{\iih a^\mu \ope{P}_\mu^{\mathrm{QM}}} \ope{X} = \ope{X}
\qquad
\ope{P}_\mu^{\mathrm{QM}} (\ope{X}) = 0
\qquad
\frac{\pd\ope{X}}{\pd x^\mu} = 0 .
	\end{equation}
Then~\eref{6.9} and~\eref{6.5} imply respectively
	\begin{gather}	\label{6.13}
\ope{P}_\mu^{\mathrm{t}}(\ope{X}) = p_\mu \ope{X}
\\			\label{6.12new1}
\ope{P}_\mu (\ope{X}) = (p_\mu+q_\mu) \ope{X} .
	\end{gather}
So, any state vector is an eigenvector for $\ope{P}_\mu^{\mathrm{t}}$
with eigenvalue $p_\mu$.
As we would like to interpret $\ope{P}_\mu^{\mathrm{t}}$ a (total)
4\ndash momentum operator of a system,~\eref{6.13} entails that $p_\mu$
\emph{should be considered as components of the total 4\ndash energy\ndash
momentum} vector $q$ of the system under consideration. Notice, the 4\ndash
vector field $p_\mu$ generally depends on the state at which the system is,
in particular $p_\mu=0$ corresponds to its vacuum state(s). Of course, the
proposed interpretation of $p_\mu$ is physically sensible if $p_\mu$ are
\emph{real} which we assume from now on.%
\footnote{~%
Since $\pd/\pd x^\mu$ is anti-Hermitian operator, the assumption that $p_\mu$
are real, which is equivalent to the Hermiticity of $p_\mu\id_\Hil$, is
tantamount to the Hermiticity of $\ope{P}_\mu^{\mathrm{t}}$,
$(\ope{P}_\mu^{\mathrm{t}})^\dagger=\ope{P}_\mu^{\mathrm{t}}$,
and, consequently, to the unitarity of
$\exp(\pm\iih a^\mu\ope{P}_\mu^{\mathrm{t}} )$,
as one expects it to be. (Recall, until now nobody has put the Hermiticity
of the momentum operator under question for any on of its definitions.)%
}
This interpretation of the numbers $p_\mu$ allows their identification with
the eigenvalues $p_\mu+q_\mu$ of the canonical momentum operator
$\ope{P}_\mu$ when it acts on state vectors, \viz we should have
	\begin{gather}	\label{6.12new2}
q_\mu = 0
\intertext{or}
			\label{6.13new}
\ope{P}_\mu (\ope{X}) = p_\mu \ope{X}
	\end{gather}
for any state vector $\ope{X}$ describing system's state with total
4-momentum vector $p_\mu$. Consequently, from~\eref{6.5}, \eref{6.9},
and~\eref{6.12new2}, we see that
	\begin{gather}	\label{6.13new1}
\ope{P}_\mu \big|_{\ope{D}_p}
=
( - \ope{P}_\mu^{\mathrm{QM}} + p_\mu\id_\Hil ) \big|_{\ope{D}_p}
\qquad
\bigl( =\ope{P}_\mu^{\mathrm{t}} |_{\ope{D}_p} \bigr) ,
\\ \intertext{where}	\label{6.13new3}
\ope{D}_p := \{ \ope{X}\in\Hil :
		\ope{P}_\mu (\ope{X}) = p_\mu \ope{X} \} ,
	\end{gather}
which, as it is easily seen, is a simple consequence of the conservation of
the energy\ndash momentum for a closed (translation invariant) system.%
\footnote{~%
It is worth noting, similar considerations in quantum mechanics give
rise to the Schr\"odinger equation.
Indeed, defining the `mathematical' energy by
$\ope{E}^{\mathrm{m}}:=\ih\frac{\pd}{\pd t}$, $x^0=:ct$, and
the `canonical' one by $\ope{E}^{\mathrm{c}}=\Ham$, $\Ham$
being the system's Hamiltonian, we see that the equation
\(
\ope{E}^{\mathrm{c}} (\psi)
=
\ope{E}^{\mathrm{m}} (\psi)
\)
is identical with the Schr\"odinger equation  for a wavefunction $\psi$. If
the system is closed, the common eigenvalues of
$\ope{E}^{\mathrm{c}}$ and
$\ope{E}^{\mathrm{m}}$
represent the (stationary) energy levels of the system under consideration.%
}
If a base of $\Hil$ can be formed from vectors in $\ope{D}_p$,
from~\eref{6.13new1} the equality~\eref{6.12} will follow.%
\footnote{~%
In Sect.~\ref{Conclusion} it will be proved that, generally, this is not the
case; see, e.g.,~\eref{C.9} and the sentence after it.%
}
But, generally, we can only assert that
	\begin{equation}	\label{6.13new2}
[ \varphi_i(x) , \ope{P}_\mu^{\mathrm{t}} ]_{\_}
=
[ \varphi_i(x) , \ope{P}_\mu ]_{\_}
=
\ih\frac{\pd\varphi_i(x) }{\pd x^\mu}
\qquad
\bigl( = - [ \varphi_i(x) , \ope{P}_\mu^{\mathrm{QM}} ]_{\_} \bigr)
	\end{equation}
which does \emph{not} imply~\eref{6.12}. The above discussion also reveals
that the equality
$\ope{P}_\mu^{\mathrm{t}}(\ope{X}) = - \ope{P}_\mu^{\mathrm{QM}}(\ope{X})$
can be valid only for states with zero 4\ndash momentum, $p_\mu=0$, \ie only
for the vacuum state(s).

	It should be mentioned, the equality~\eref{6.12} entails~\eref{6.4}
for arbitrary operator $\ope{A}(x)$, which, as pointed in
footnote~\ref{CounterExample}, leads to contradictions.

	From~\eref{6.13} or directly from the explicit relation~\eref{6.9},
we derive
	\begin{equation}	\label{6.14}
\e^{-\iih a^\mu\ope{P}_\mu^{\mathrm{t}}} (\ope{X})
=
\e^{\iu ( \frac{1}{\hbar} a^\mu p_\mu) } \ope{X} .
	\end{equation}
Hence, the action of the unitary operators
$U(a,\bs 1):=\e^{-\iih a^\mu\ope{P}_\mu^{\mathrm{t}}}$,
which form a unitary representation of the translation subgroup of the
Poincar\'e group, on state vectors reduce to a simple multiplication
with a phase factor. This means that the vectors $\ope{X}$ and
$U(a,\bs1)(\ope{X})$ describe one and the same state of the system as the state
vectors, in general, are defined up to a phase factor.

	As we see, the situation with $\ope{P}_\mu^{\mathrm{t}}$ is
completely different from the one with $\ope{P}_\mu^{\mathrm{QM}}$ for which
$\ope{P}_\mu^{\mathrm{QM}} (\ope{X}) \equiv 0$, \ie if one takes
$\ope{P}_\mu^{\mathrm{QM}}$ as a `true' momentum operator, any state will be
characterized by identically vanishing 4\ndash momentum vector.

	The relation
	\begin{equation}	\label{6.14new}
\ope{A}(x+a)
=
\e^{-\iih a^\mu\ope{P}_\mu^{\mathrm{t}}}
\circ\ope{A}(x)
\circ \e^{+\iih a^\mu\ope{P}_\mu^{\mathrm{t}}}
	\end{equation}
is a corollary of~\eref{6.10} and~\eref{6.2} as $p_\mu\id_\Hil$,
$p_\mu=\const$, commutes with all operators on $\Hil$. Hence,
$\ope{P}_\mu^{\mathrm{t}}$, as given by~\eref{6.9}, is a generator of (the
representation of) the translations on the operators on $\Hil$. But, in view
of~\eref{6.14}, $\ope{P}_\mu$ is \emph{not} a generator of (the
representation of) the translations on the vectors in $\Hil$. As a result of
~\eref{6.13new1}, the same conclusions are valid and with respect to the
canonical momentum operator $\ope{P}_\mu$ on the domain $\ope{D}_p$, defined
via~\eref{6.13new3}, for any given $p$.

	Ending this section, we want to note that the components of momentum
operator(s) commute. In fact, equation~\eref{6.4} with
$\ope{A}(x)=\ope{P}_\nu$ implies
	\begin{equation}	\label{2.15}
[ \ope{P}_\mu,\ope{P}_\nu ]_{\_} =0
	\end{equation}
due to~\eref{6.1new}.%
\footnote{~%
This proof of~\eref{2.15} is not quite rigorous as, in view of~\eref{6.1},
$\ope{P}_\mu$ is not a function, but an operator\ndash valued functional of
the field operators $\varphi_i(x)$. Rigorously~\eref{2.15} is a corollary
of~\eref{6.1}, the conservation law~\eref{6.1new}, and the equal\ndash time
(anti-)commutation relations~\cite{Roman-QFT,Bjorken&Drell}. For other proof,
see, e.g.,~\cite[p.~76]{Roman-QFT}.%
}
Similar equality for the translation momentum operator
$\ope{P}_\mu^{\mathrm{t}}$, \ie
	\begin{equation}	\label{2.16}
[ \ope{P}_\mu^{\mathrm{t}},\ope{P}_\nu^{\mathrm{t}} ]_{\_} =0
	\end{equation}
is a direct consequence of~\eref{6.10} due to $\pd_\nu p_\mu\equiv0$ (by
definition $p_\mu$ are constant (real) numbers).


\section {Conclusion}
\label{Conclusion}

	The main results of our previous exposition can be formulated as
follows.
	\begin{description}
\item[(i)]
	The generator $\ope{P}_\mu^{\mathrm{QM}}$ of (the representation of)
the translations in system's Hilbert space is \emph{not} the momentum
operator in quantum field theory. However, there is a close connection between
both operators (see equation~\eref{6.9}).

\item[(ii)]
	The translation momentum operator $\ope{P}_\mu^{\mathrm{t}}$ of a
quantum system is a generator of (the representation of) the translations in
the space of operators acting on system's Hilbert space. It depends on a
4\ndash vector $p$ with constant (relative to Poincar\'e group) components.

\item[(iii)]
	The (canonical/physical) momentum operator $\ope{P}_\mu$  is,
generally, different from the translation momentum operator. However, the
restrictions of $\ope{P}_\mu$ and $\ope{P}_\mu^{\mathrm{t}}$ on the
set~\eref{6.13new3} coincide due to the identification of the vector  $p$
with the vector of eigenvalues of $\ope{P}_\mu$.

\item[(iv)]
	When commutators with field operators or functions of them are
concerned, the operators $\ope{P}_\mu$ and $\ope{P}_\mu^{\mathrm{t}}$ are
interchangeable (see~\eref{6.13new2}). However, equalities, like~\eref{6.2'},
in particular~\eref{6.7}, are identities, while ones, like~\eref{6.8}, are
equations relative to the field operators and their validity depends on the
particular Lagrangian from which $\ope{P}_\mu$ is constructed.

%
	\end{description}

	As it is noted in~\cite[\S~68]{Bjorken&Drell-2}, the quantum field
theory must be such that the (canonical) momentum operator $\ope{P}_\mu$,
given in Heisenberg picture via~\eref{6.1}, must satisfy the Heisenberg
relations/equations~\eref{6.8}. This puts some restrictions on the
arbitrariness of the (canonical) energy\ndash momentum tensorial operator
$\ope{T}^{\mu\nu}$ entering into~\eref{6.1} and obtained, via the (first)
Noether theorem, from the system's Lagrangian. Consequently, at the end,
this puts some, quite general, restrictions on the possible Lagrangians
describing systems of quantum fields.

	Our analysis of the momentum operator in quantum field theory can be
transferred \emph{mutatis mutandis} on the angular momentum operator and
similar ones arising via the (first) Noether theorem from  Lagrangians
invariant under some spacetime continuous symmetries.

	Since in the description of the dynamics of a quantum system enters
only the (physical) momentum operator $\ope{P}_\mu$, we share the opinion
that it is more important than the mathematical momentum operator
$\ope{P}_\mu^{\mathrm{t}}$; the latter one playing an auxiliary role mainly
in the derivation of Heisenberg equations of motion or the transformation
properties of quantum fields.

	Now we would like to look on the non-relativistic and relativistic
quantum mechanics from the view\ndash point of the above results. Since these
theories are, usually, formulated in the Schr\"odinger picture of motion, we
shall, first of all, `translate' the momentum operator into it. Besides, as in
quantum field theory only Hamiltonians, which do not explicitly depend on the
spacetime coordinates are considered, we shall suppose the system's Hamiltonian
 $\Ham=c\ope{P}_0$ to be of such a type in Heisenberg picture of motion.

	The transition from Heisenberg picture to Schr\"odinger one is
performed via the mappings
	\begin{align}	\label{C.1}
\ope{X}\mapsto & \lindexrm[\mspace{-6mu}\ope{X}]{}{S} = \e^{\iih(t-t_0)\Ham}
\\			\label{C.2}
\ope{A}(x)\mapsto & \lindex[\mspace{-8mu}\ope{A}]{}{S}(x)
=
 \e^{\iih(t-t_0)\Ham}  \circ\ope{A}(x)\circ   \e^{-\iih(t-t_0)\Ham} ,
	\end{align}
where $\ope{X}\in\Hil$ and $\ope{A}(x)\colon\Hil\to\Hil$ are arbitrary, $t$
is the time (coordinate) and $t_0$ is arbitrarily fixed instant of time. In
particular,~\eref{C.2} with $\ope{A}(x)=\ope{P}_\mu^{\mathrm{t}}$, gives
(see~\eref{6.10} and recall that $x^0=ct$, $c$ being the velocity of light in
vacuum)
	\begin{subequations}	\label{C.3}
	\begin{align}	\label{C.3a}
c \lindexrm[\mspace{-2mu}\ope{P}]{}{S}^{\mathrm{t}}_0
& =
c \ope{P}_0^{\mathrm{t}} + c \ope{P}_0
=
c \ope{P}_0^{\mathrm{t}} + \Ham
= - \ih\frac{\pd}{\pd t} + e\id_\Hil  + \Ham
\\			\label{C.3b}
\lindexrm[\mspace{-2mu}\ope{P}]{}{S}^{\mathrm{t}}_a
& =
\ope{P}_a^{\mathrm{t}}
= - \ope{P}_a^{\mathrm{QM}} +p_a\id_\Hil
= - \ih\frac{\pd}{\pd x^a} +p_a\id_\Hil
\qquad
a=1,2,3 .
	\end{align}
	\end{subequations}
Here $p_0=\frac{1}{c}e$ is an eigenvalue of
\(
\ope{P}_0 = \frac{1}{c}\Ham
= \frac{1}{c}\lindexrm[\Ham]{}{S}
=   \lindexrm[\mspace{-2mu}\ope{P}]{}{S}_0 .
\)

	Let $\lindexrm[\mspace{-6mu}\ope{X}]{}{S}_e$ be a state vector
corresponding to a state with fixed energy $e$, \ie
	\begin{equation}	\label{C.5}
\lindexrm[\mspace{0mu}\Ham]{}{S} ( \lindexrm[\mspace{-6mu}\ope{X}]{}{S}_e)
=
\Ham ( \lindexrm[\mspace{-6mu}\ope{X}]{}{S}_e)
=
e \lindexrm[\mspace{-6mu}\ope{X}]{}{S}_e
\quad
\Ham (\ope{X}_e) = e \ope{X}_e.
	\end{equation}
A straightforward calculation gives
	\begin{gather}	\label{C.6}
c\ope{P}_0^{\mathrm{t}} ( \lindexrm[\mspace{-6mu}\ope{X}]{}{S}_e )
=
- \ih\frac{\pd \lindexrm[\mspace{-6mu}\ope{X}]{}{S}_e } {\pd t}
+ e \lindexrm[\mspace{-6mu}\ope{X}]{}{S}_e
= 0
\\
	\intertext{which, in view of~\eref{C.5}, is a version of the
Schr\"odinger equation}
			\label{C.7}
\ih \frac{\pd \lindexrm[\mspace{-6mu}\ope{X}]{}{S}_e} {\pd t}
= \lindexrm[\mspace{0mu}\Ham]{}{S}
(\lindexrm[\mspace{-6mu}\ope{X}]{}{S}_e) .
\\
	\intertext{Besides, equations~\eref{C.3a}
and~\eref{C.6} imply}
			\label{C.8}
c \lindexrm[\mspace{-2mu}\ope{P}]{}{S}_0^{\mathrm{t}}
( \lindexrm[\mspace{-6mu}\ope{X}]{}{S}_e )
=
  \lindexrm[\mspace{0mu}\Ham]{}{S}
( \lindexrm[\mspace{-6mu}\ope{X}]{}{S}_e )
=
  \Ham ( \lindexrm[\mspace{-6mu}\ope{X}]{}{S}_e )
=
c \ope{P}_0 ( \lindexrm[\mspace{-6mu}\ope{X}]{}{S}_e )
= e \lindexrm[\mspace{-6mu}\ope{X}]{}{S}_e .
\\
	\intertext{Therefore, if $\ope{X}_e\not=0$, then (see~\eref{C.6})}
			\label{C.9}
\ope{P}_0^{\mathrm{t}} (\lindexrm[\mspace{-6mu}\ope{X}]{}{S}_e )
= 0
\not=
\frac{e}{c} \lindexrm[\mspace{-6mu}\ope{X}]{}{S}_e
=\ope{P}_0( \lindexrm[\mspace{-6mu}\ope{X}]{}{S}_e ) .
	\end{gather}
Hence, the non-zero state vectors, representing states with fixed and
non-vanishing energy in the Schr\"odinger picture, can serve as example of
vectors on which the equality~\eref{6.12} (in Heisenberg picture)
\emph{cannot} hold. However, it must be emphasized on the fact that a vector
 $\lindexrm[\mspace{-6mu}\ope{X}]{}{S}_e$  with
 $\lindexrm[\mspace{-6mu}\ope{X}]{}{S}_e\not=0$ does not represent a
physically realizable state in Heisenberg picture as it is a time\ndash
depending vector, contrary to the physically realizable ones which, by
definition, are constant in this picture of motion.

	Consider now states with fixed 3-momentum vector
$\bs p=(p^1,p^2,p^3)=-(p_1,p_2,p_3)$.%
\footnote{~%
The Lorentz metric is suppose to be of signature $(+\,-\,-\,-)$. Here and
below the boldface symbols denote 3\ndash dimensional vectors formed from the
corresponding 4\ndash vectors.%
}
In 3\ndash dimensional notation, the equation~\eref{C.3b} reads
	\begin{equation}	\label{C.10}
\lindexrm[\mspace{-2mu}\bs{\ope{P}}]{}{S}^{\mathrm{t}}
=
\bs{\ope{P}}^{\mathrm{t}}
=
- \lindexrm[\mspace{-2mu}\bs{\ope{P}}]{}{S}^{\mathrm{QM}} + \bs{p}\id_\Hil
=
 \ih \bs\nabla + \bs p\id_\Hil
	\end{equation}
with
\(
\bs\nabla
:=
\bigl( \frac{\pd}{\pd x^1},\frac{\pd}{\pd x^2},\frac{\pd}{\pd x^3} \bigr)
\)
 and $\bs p$ being an eigenvector of $\bs{\ope{P}}$.
Suppose $\ope{X}_{\bs p}$  is an eigenvector of $\bs{\ope{P}}$ with $\bs p$
as eigenvalue, \viz
	\begin{equation}	\label{C.11}
\bs{\ope{P}} (\ope{X}_{\bs p}) = \bs p  \ope{X}_{\bs p}
\qquad
\lindexrm[\mspace{-2mu}\bs{\ope{P}}]{}{S}
	     ( \lindexrm[\mspace{-6mu}\ope{X}]{}{S} _{\bs p} )
= \bs p \,  \lindexrm[\mspace{-6mu}\ope{X}]{}{S} _{\bs p}  .
	\end{equation}

	Since in non-relativistic quantum mechanics is assumed
$ \lindexrm[\mspace{-6mu}\ope{X}]{}{S} _{\bs p} $
to be also an eigenvector of
\(
\bs{\ope{P}}^{\mathrm{QM}}
=
 - \ih\bs\nabla
=
\lindexrm[\mspace{-2mu}\bs{\ope{P}}]{}{S} ^{\mathrm{QM}}
\)
with the same eigenvalues, \ie
	\begin{equation}	\label{C.12}
\lindexrm[\mspace{-2mu}\bs{\ope{P}}]{}{S} ^{\mathrm{QM}}
	( \lindexrm[\mspace{-6mu}\ope{X}]{}{S} _{\bs p} )
=
\bs p \, \lindexrm[\mspace{-6mu}\ope{X}]{}{S} _{\bs p}
	\end{equation}
or, equivalently,
\(
\lindexrm[\mspace{-6mu}\ope{X}]{}{S} _{\bs p}
=
\e^{-\iih(\bs x-\bs x_0)\cdot\bs p}
	( \lindexrm[\mspace{-6mu}\ope{X}]{}{S} _{\bs p} |_{x_0} ) ,
\)
equation~\eref{C.10} yields
	\begin{equation}	\label{C.13}
\lindexrm[\mspace{-2mu}\bs{\ope{P}}]{}{S} ^{\mathrm{t}}
	( \lindexrm[\mspace{-6mu}\ope{X}]{}{S} _{\bs p} )
=
\bs{\ope{P}}^{\mathrm{t}} ( \lindexrm[\mspace{-6mu}\ope{X}]{}{S} _{\bs p} )
\equiv 0
	\end{equation}
which in the Heisenberg picture reads
	\begin{equation}	\label{C.14}
\bs{\ope{P}}^{\mathrm{t}}  ( \ope{X}_{\bs p} ) = 0 .
	\end{equation}
	By virtue of~\eref{C.11}, the last equality means that
	\begin{equation}	\label{C.15}
\bs p = 0
	\end{equation}
if $\ope{X}_{\bs p}\not=0$.
Consequently, from the view-point of quantum field theory, the quantum
mechanics describes systems with zero 3\ndash momentum. This unpleasant
conclusion is not rigorous. It simply shows that~\eref{C.12} is not
compatible with other axioms of quantum field theory or, said differently,
the quantum mechanics and quantum field theories rest on different, not
completely compatible postulates. Since we plan to give a satisfactory
solution of the problem of comparison of the grounds of and interrelations
between these theories elsewhere, below are presented non\ndash rigorous,
possibly intuitive and naive, conclusions from the above\ndash written
material.

	The non-relativistic quantum mechanics can be obtained from quantum
field theory in Schr\"odinger picture by extracting from the latter theory
only the Schr\"odinger equation~\eref{C.7} and ignoring all other aspects of
it. Besides, this equation is generalized in a sense that it is assumed to
hold for arbitrary, generally non\ndash closed or translation
non\ndash invariant, systems, \ie
	\begin{equation}	\label{C.16}
\ih \frac{\pd \lindexrm[\mspace{-6mu}\ope{X}]{}{S}}{\pd t}
=
\lindexrm[\mspace{0mu}\ope{H}]{}{S} (t,\bs x)
( \lindexrm[\mspace{-6mu}\ope{X}]{}{S} ) .
	\end{equation}
The explicit form of the Hamiltonian
$\lindexrm[\mspace{0mu}\ope{H}]{}{S} (t,\bs x)$,
as well as of other operators, representing dynamical variables,  are almost
`put by hands' by postulating them; the only guiding principle being the
compatibility with some classical analogues, if such exist for a given
variable or system. In particular, it happens that the operator representing
the 3\ndash momentum vector is exactly
$\bs{\ope{P}}^{\mathrm{QM}}=-\ih\bs\nabla$, \ie
\(
\ope{P}^{\mathrm{QM}}_a
=
- ( \bs{\ope{P}}^{\mathrm{QM}}) ^a
=
\ih \frac{\pd}{\pd x^a} ,
\)
 $a=1,2,3$.

	Now a few words about relativistic quantum mechanics are in order.
The situation in that case is similar to the non\ndash relativistic one with
the only difference that the Hamiltonian operator
$\lindexrm[\mspace{0mu}\ope{H}]{}{S} (t,\bs x)$ should be consistent with
special relativity, not with classical mechanics. For instance, for a point
particle with rest mass $m$, it must be such that
	\begin{equation}	\label{C.17}
( \lindexrm[\mspace{0mu}\ope{H}]{}{S} (t,\bs x) )^2
=
c^2 (\bs{\ope{P}}^{\mathrm{QM}})^2 + m^2c^4 \id_\Hil
	\end{equation}
where  $(\bs{\ope{P}}^{\mathrm{QM}})^2$ is the square of the 3\ndash
dimensional part of the operator~\eref{6.6}. From this relation, under some
additional assumptions, the whole relativistic quantum mechanics can be
derived, as it is done, e.g., in~\cite{Bjorken&Drell-1}.


\addcontentsline{toc}{section}{References}
\bibliography{bozhopub,bozhoref}
\bibliographystyle{unsrt}
\addcontentsline{toc}{subsubsection}{This article ends at page}

\end{document}